\documentclass[USenglish,twocolumn]{article}

\usepackage[utf8]{inputenc}
\usepackage[big]{dgruyter}
\usepackage{listings}
\usepackage{graphicx}

\begin{document}

  \articletype{Research Article{\hfill}}

  \author*[1]{Tosin P. Adewumi}

\author[2]{Marcus Liwicki}

  \affil[1]{Lule\aa \ University of Technology, E-mail: oluwatosin.adewumi@ltu.se}

  \affil[2]{Lule\aa \ University of Technology, E-mail: marcus.liwicki@ltu.se}

  \title{\huge Inner For-Loop for Speeding Up Blockchain Mining}

  \runningtitle{Inner For-Loop for Speeding Up Blockchain Mining}


  \begin{abstract}
{In this paper, the authors propose to increase the efficiency of blockchain mining by using a population-based approach. Blockchain relies on solving difficult mathematical problems as proof-of-work within a network before blocks are added to the chain.
Brute force approach, advocated by some as the fastest algorithm for solving partial hash collisions and implemented in Bitcoin blockchain, implies exhaustive, sequential search.
It involves incrementing the nonce (number) of the header by one, then taking a double SHA-256 hash at each instance and comparing it with a target value to ascertain if lower than that target.
It excessively consumes both time and power. In this paper, the authors, therefore, suggest using an inner for-loop for the population-based approach. Comparison shows that it's a slightly faster approach than brute force, with an average speed advantage of about 1.67\% or 3,420 iterations per second and 73\% of the time performing better.
Also, we observed that the more the total particles deployed, the better the performance until a pivotal point. Furthermore, a recommendation on taming the excessive use of power by networks, like Bitcoin's, by using penalty by consensus is suggested.}
\end{abstract}

  \journalname{Open Comput. Sci}

\DOI{https://doi.org/10.1515/comp-2020-0004}
  \startpage{42}
  \received{June 07, 2019}
  \accepted{Nov 18, 2019}

  \journalyear{2020}
  \journalvolume{10}
  \journalissue{}

\maketitle
\section{Introduction}
Blockchain, as a digital distributed ledger, implies it’s an electronic, decentralized permanent storage of transactions \cite{nakamoto2009}. Its first successful implementation was as the backbone of a peer to peer network for Bitcoin by Nakamoto \cite{nakamoto2009}. It has no requirement for a central server. Instead, it's designed to operate within a network on a consensus basis \cite{narayanan2016bitcoin}. Blockchain is one of the latest and, arguably, the most promising technology for virtual money, given the other previous attempts that have failed \cite{narayanan2016bitcoin}. It has potential usage in many fields and industries beyond money-related instances, for example, in smart contracts, smart properties, legal documents, health records, marriage certificates and a number of others \cite{crosby2016blockchain}.\\

The research question being addressed here is 'How can a population-based algorithm be applied to speed up blockchain mining, if at all it can?' This is because it has been established in other literature that optimization algorithms or population-based searches work well for certain problems and poorly for others \cite{clerc2011theory, settles2005introduction}. Lehre and Witt \cite{lehre2013finite} showed that standard or basic particle swarm optimization (PSO) with one or two particles stagnates even on one-dimensional sphere. Population-based approaches use multiple solutions at the same time \cite{Beheshti2013}, such as with genetic algorithms \cite{holland1992genetic, Beheshti2013} or PSO \cite{eberhart1995new}. Brute force is the widespread implementation, possibly the only approach, to mining blockchain \cite{narayanan2016bitcoin, nakamoto2009}. This research uses the method of comparative study to critically analyze results from several tests of the implemented population-based algorithm and sequential brute force algorithm.
The population-based algorithm is a simple case of inner for-loop (IFL) program construct, i.e., a nested for-loop within another loop.
Brute force approach usually uses one loop to iterate over all values and is the most intuitive approach to solving a problem but has very low efficiency in some cases \cite{willis2004power}.\\

The main contribution of this research is the implementation of the inner for-loop algorithm, demonstrated by Adewumi \cite{adewumi2018inner}, to mine blockchain faster.
The relevance of this to power/energy consumption of the network machines will be mentioned briefly in the discussion section of this paper.
In addition, a recommended total number of particles for the population is determined.
The search algorithm may be relevant and applicable to other search scenarios that are reliant on brute force, for example, cryptography. 
The scope of this study will include inner for-loop construct, blockchain network and SHA-256 hash. This research does not address application-specific integrated circuit-resistance (ASIC-resistance).
The sections that follow include a brief literature review, methodology, results, discussion and conclusion.\\

\section{Literature Review}
According to Narayanan et al \cite{narayanan2016bitcoin}, the concept of blockchain can be traced back years ago to Haber and Stornetta (1991), who proposed secure timestamping for digital documents. An anonymous individual or group, using the name Satoshi Nakamoto, published 'Bitcoin: A Peer-to-Peer Electronic Cash System' in 2008 \cite{nakamoto2009,crosby2016blockchain}. Nakamoto \cite{nakamoto2009} set out to solve the trust issue with regards to electronic transactions through the creation of blockchain, the decentralized, technological foundation of Bitcoin cryptocurrency, consequently eliminating third-party financial institutions. In the network, a new transaction is broadcast to all miners (nodes or machines) and each collects it and begins to find the mathematical solution (called proof-of-work), by hashing values so that the result begins with a number of zero bits. This is such that it falls below the periodically reviewed target set by the network system. The proof-of-work (PoW) is verifiable by executing a single hash. \cite{nakamoto2009}\\

More specifically, sequential, brute force search is carried out by incrementing the nonce (number) of a block header by one, then taking a double SHA-256 hash at each instance and comparing it with a system-determined target value to ascertain if lower than that target \cite{back2002hashcash, nakamoto2009}.
Each header is made up of a previous hash, merkle root, version, timestamp, bits (indicating the difficulty level or target) and a nonce \cite{nakamoto2009, narayanan2016bitcoin}.
The node that finds the solution creates a new block based on it and broadcasts the new block to all other nodes, who then verify if the proof-of-work is correct and the value has not been spent.
The nodes then add the new block to their chain chronologically and prepare to create a new block by using the hash of the accepted block as the previous block \cite{nakamoto2009}.
This difficult proof-of-work, which consumes time and excessive power, has led to the production of special-purpose processors called application-specific integrated circuits (ASICs) for blockchain mining \cite{narayanan2016bitcoin}.
However, Bitcoin's network works by adjusting the difficulty of the mathematical PoW so that it takes an average of ten minutes to solve one \cite{narayanan2016bitcoin}.
Hence, the difficulty in its network has been growing steadily over the years because of the powerful ASICs added to, or used to replace older nodes. \cite{narayanan2016bitcoin}.\\

Adam Back's \cite{back2002hashcash} hashcash proof-of-work model was used in the implementation of a distributed timestamp on a peer-to-peer basis in Bitcoin \cite{nakamoto2009}.
The longest chain in the network, always considered the correct one, will have the greatest proof-of-work invested and represents the majority decision. If the PoW is generated too fast, the difficulty increases and this is usually done every two weeks, based on the average ten minutes the network expects to find a solution each time \cite{nakamoto2009, narayanan2016bitcoin}. An attacker will have to do the proof-of-work of a past chain and all previous chains to the point of the genesis block to be able to modify a past block, which is near unrealistic.
It has been asserted that the fastest algorithm for computing partial hash collisions, upon which hashcash cost-function is based, is brute force \cite{back2002hashcash}. \\

The nonce, timestamp and bits fields are 32 bits long.
In generating a valid nonce that solves the mathematical header problem, after all possible values have been exhausted in the 32-bit nonce and no valid hash found, the value in the coinbase can be modified, thereby triggering changes in the merkle tree, and the process of iterating through possible values of nonce repeated.
Other fields that can be modified in the header to repeat the iteration through the nonce when no valid hash is found are the timestamp and merkle root \cite{narayanan2016bitcoin}.
The timestamp is registered in epoch time, that is, the time in seconds from 1970 UTC.
Mining difficulty changes every 2,016 blocks, as a step function, which translates to about every two weeks.\\

Secure Hash Algorithm-256 (SHA-256) \cite{pub2012secure} is a keyless, cryptographic hash function with fixed output of 256 bits created by the National Institute of Standards and Technology (NIST) in 2002 alongside two other versions, based on previous standards \cite{sobti2012cryptographic}.
Hash functions take in input string of any size and efficiently compute output string of fixed size called digest \cite{pub2012secure,narayanan2016bitcoin}.
Collision, which implies two different inputs having the same hashed output, is actually possible in hash functions, especially considering the birthday paradox in probability \cite{narayanan2016bitcoin}.
This is, however, not practical to compute because of the astronomically long time it will take computers to do so \cite{narayanan2016bitcoin, sobti2012cryptographic}.
There are possible attacks on cryptographic hash functions, like pre-image attack or collision attack, however, there hasn’t been any reported successful attack on SHA-256 \cite{sobti2012cryptographic}.
The ever-present risk of possible attacks with previous cryptographic functions led NIST to organize competitions that led to the creation of the SHA-3 family of functions \cite{dworkin2015sha}.
Compared to SHA-3, however, SHA-256 is still more popular.
This is because support for SHA-1 has been discontinued, SHA-3 is relatively new and SHA-256 has been the secure platform for internet and network communication \cite{pub2012secure,dworkin2015sha}.\\ 

In order to address the huge advantage ASICs have over regular computers, some researchers have attempted new PoW algorithms that introduce ASIC-resistance, like Scrypt \cite{raikwar2019sok}, Ethash \cite{wood2014ethereum, zamanov2018asic} and Equihash \cite{biryukov2017equihash}.
These algorithms generally, though unsuccessfully over time \cite{biryukov2017equihash, raikwar2019sok}, try to place restrictions on the resources of ASICs  by being memory-hard in some cases.
Bitmain, an ASIC manufacturer, overturned the objective of such algorithms by producing ASICs optimized over such algorithms \cite{raikwar2019sok}.
Equihash family of proofs of work also uses parts of the algorithm employed in Bitcoin, particularly a fixed-time difficulty filter requiring a nonce and leading zeros in hash.
This is besides additional, modified Wagner's algorithm.
Equihash is not claimed to perform faster than brute force, with regards to Bitcoin.
Indeed, Biryukov and Khovratovich, who introduced Equihash, pointed out that the best algorithms for NP problems 'run in exponential time' \cite{biryukov2017equihash}, as do Wagner's algorithm and brute force.
\\

\section{Methodology}
This research adopted a comparative study method in order to evaluate brute force algorithm and the newly implemented inner for-loop algorithm, using various number of total particles in order to ascertain the preferred total for the system. Although power analysis to determine the sample size to use from the blockchain population in order to compare two means showed that a higher sample size is desirable for greater power for small effect size, 120 was tested. The basic data were obtained by secondary means from the website that keeps such records (www.blockchain.com). These include values of block header fields of certain blocks in the Bitcoin blockchain. Simple random sampling was used to pick all blocks by generating the blocks to be picked using a simple Python program that randomly picked from 1 to 557,132 blocks available as at January 2019 \cite{kazmier2004theory}.\\

The programs were written in Python (version 3.7) programming language, using an HP desktop having Intel Core i5-2500 at 3.30Ghz clock speed, 8GB RAM and 64bit Windows 10 operating system.
Python, apparently, was more suitable as a programming language because of its int datatype’s unrestricted size \cite{swaroop2013byte}, which is required when dealing with the huge numbers blockchain operates with. For IFL, the maximum iteration value for the outer loop is determined by capping it as the result of dividing the highest possible value of 2 to power 32 (since this is the highest possible value for the nonce field) by the particle total. If this maximum iteration value is left simply as 2 to power 32 as with the brute force program, then the efficiency of the program will be lost in cases when a hash solution or valid nonce is not found and adjustment needs to be made to the header before restarting the search process. The code is provided in the appendix for scrutiny and easy replication of this study.\\

Since memory access is slower for variables than having constants, the result of the earlier cap can be, and was imputed, as a constant value for the outer while loop for each particle total. The number of hash attempts for brute force cases are the same for the IFL approach. It was ensured no other unrelated application was open and running while each program executed. These precautions minimized threats to internal validity so that things were measured right. Parallel programming or multiple threads were not employed in the codes, neither was machine-dependent optimization used.\\

The IFL program was run for each of the solved Bitcoin blockchain headers using several particle totals of 2, 6, 20, 100, 200 and 1000. Due to the extensive length of time running each test from nonce iteration of 0 to the end will take, each was given a cap of 30,000,000 iterations and the execution time recorded. Two runs were carried out per particle total per block and average time calculated. The iterations per second was then calculated from that, which was used to obtain time to valid nonce for each block. This resulted in a total of 1,680 runs, including for brute force. Three blocks had nonce within the 30,000,000 iterations and were correctly returned during the tests.

\section{Results}
\begin{figure}[h]
    \centering
    \includegraphics[width=0.5\textwidth]{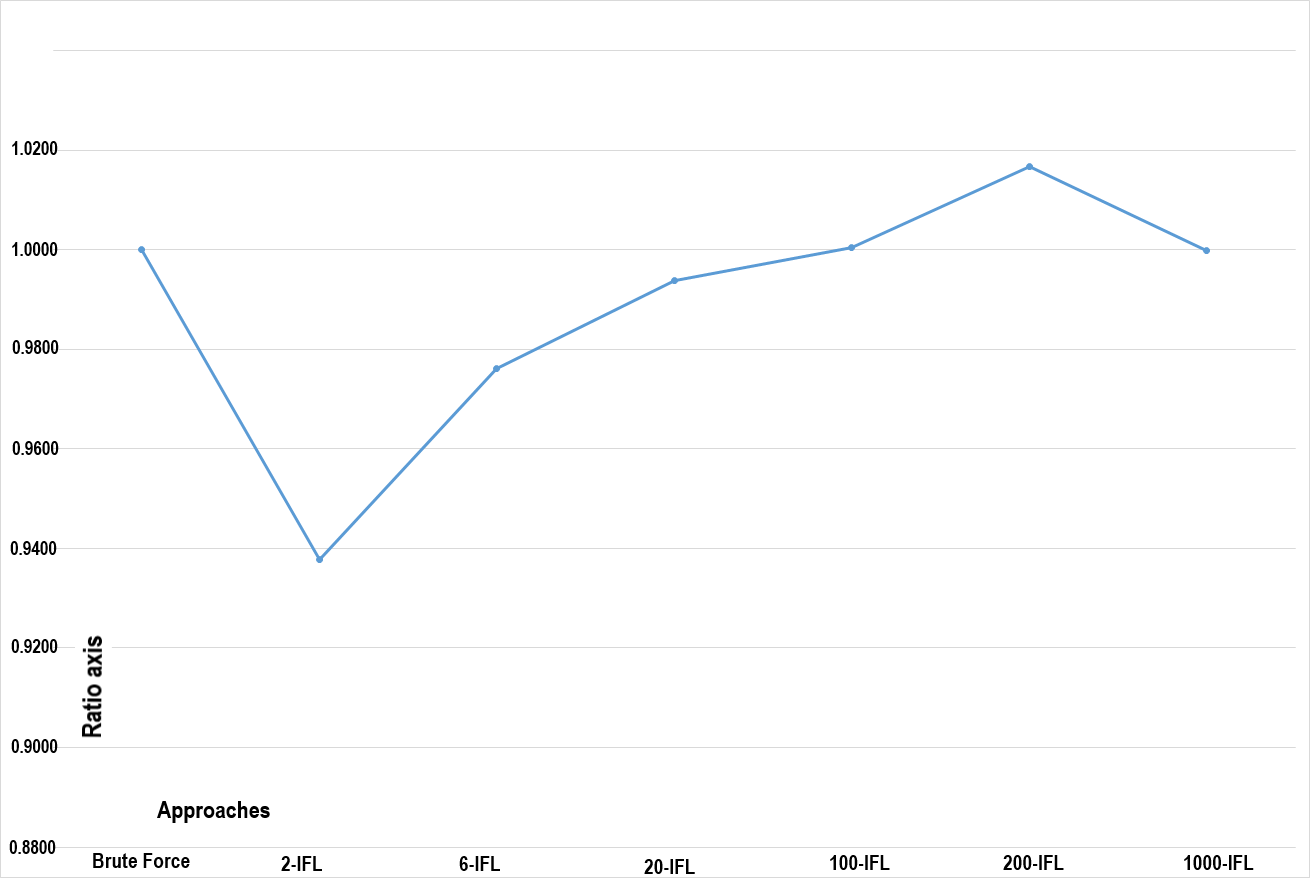}
    \caption{Speed Ratio Comparison of Approaches}
    \label{fig:my_label}
\end{figure}

\begin{figure}[h]
    \centering
    \includegraphics[width=0.5\textwidth]{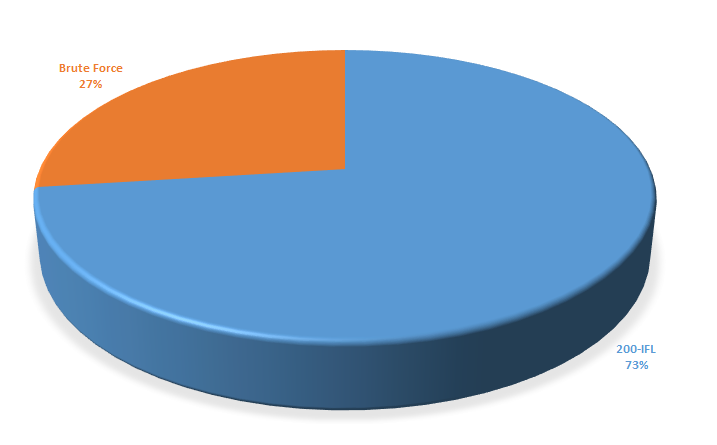}
    \caption{Better Performance Chart}
    \label{fig:pie_c}
\end{figure}

After 1,680 experimental runs, results were obtained and tabulated. It was observed that the 200-Particle total is the best of the approaches tried. It had the lowest time minimization though the speed advantage over brute force was not very much. The 100-Particle IFL case is also better than brute force, though lesser than its 200-Particle counterpart. The average speed improvement of IFL (with 200 particles) over BF is 1.67\% (about 3,420 iterations per second) and 73\% of the test instances show IFL performing better. Performance from 2-Particle total starts out poorly, but as the number of particles increases, the speed improves and time minimization gets better until after 200 particles. Table~\ref{tab1} gives a summary of the comparison of the key indicators of the various approaches and this is expressed pictorially in Figure \ref{fig:my_label}, while Figure \ref{fig:pie_c} shows how often brute force and 200-IFL performed better than each other. Our null hypothesis that there is no difference between the means of brute force and 200-Particle IFL is tested with statistical analysis of two sample t-test. It shows that the computed \textit{t} had a value of 5.953 against the critical value of 2.576 (given alpha is 0.01), hence, we reject the null hypothesis as there is a significant difference between the two means.\\

\begin{table}[h]
\caption{Comparison of Approaches}
\begin{center}
\begin{tabular}{l|c|c}
\hline
\textbf{Approach} & \textbf{Mean iterations/s} & Ratio to BF \\
\hline
Brute Force (BF) &  204,970 & 1.0000 \\
\hline
2-Particle IFL & 192,198 & 0.9377 \\
\hline
6-Particle IFL &  200,060 & 0.9760 \\
\hline
20-Particle IFL &  203,708 & 0.9938 \\
\hline
100-Particle IFL &  205,074 & 1.0005 \\
\hline
200-Particle IFL & 208,393 & \textbf{1.0167} \\
\hline
1000-Particle IFL &  204,932 & 0.9998 \\
\hline
\end{tabular}
\label{tab1}
\end{center}
\end{table}

\section{Discussion}
The results suggest the new algorithm utilized the machine-independent optimization noted with IFL \cite{adewumi2018inner}. Further work might show that this advantage may also be noticeable on a multiplied scale when parallel programming and multiple threads are employed.\\

Given the fact that Bitcoin network's self-adjusting average mining time of 10 minutes has not stopped enthusiasts from securing more powerful machines to mine even faster for the reward it offers \cite{narayanan2016bitcoin, crosby2016blockchain}, then utilizing a machine-independent algorithm, such as IFL, can also give an advantage.
This is with the knowledge that the system will still adjust its difficulty in the long run. Besides, it would save some energy if it takes lesser time to solve the proof-of-work. It would, therefore, be desirable to have a faster approach.
One solution to the ever-growing demand for more power in a blockchain network, such as Bitcoin's, is to introduce \textbf{penalty by consensus} into such a network.
This concept of penalty by consensus, independent of IFL, can be used to disqualify nodes which go above a dynamic power threshold set by the consensus of the network.
A future work can compare this new concept to ASIC-resistant algorithms mentioned earlier and energy measurements evaluated.\\

\section{Conclusion}
This research work used the method of comparative study to critically analyze results from several scenarios of the newly implemented inner for-loop program and the widespread brute force algorithm on blockchain mining. From the results obtained and analysis carried out, we can observe that this population-based algorithm (implemented as a simple inner for-loop) can successfully be applied to speed up blockchain mining. Furthermore, we observed that fewer particles gave poor result but the more particles were deployed in a population, the better the general performance until a pivotal point. A recommended total number of particles for the population was determined as 200 out of the several tried. Further work can be carried out with a larger sample size. Other particle totals beyond 1000 may be experimented with to find out if more optimal solutions exist. The inner for-loop algorithm may also be tested on parallel processors. Different programming languages may also be implemented to compare with the python implementation for further external validity. Another implication of this research worth investigating is the security of current cryptography algorithms popularly implemented in brute force.

\bibliographystyle{unsrt}
\bibliography{bibfile}

\section*{Appendix}
\begin{lstlisting}[language=Python, caption=Inner For Loop Example with 200 Particles, breaklines]
import time
import math
from hashlib import sha256
from struct import pack
from binascii import hexlify
from codecs import decode

version = pack("<L", 0x01)
previous_header_hash = decode("000000000009852382d88a5442c35c60214
0bf08a6c40f9d9475326572032ecf", 'hex')[::-1]
merkle_root = decode("34d96742a37b46acd0d52a19be1bc0a2c4
50b3c94595117ce82dd03b39930570", 'hex')[::-1]
date = pack("<L", 1290624043)
nbits = pack("<L", 453610282)  	
nbits_calc = hexlify(nbits[::-1])
base = 256
exponent = int(nbits_calc[0:2], 16) - 3
significand = int(nbits_calc[2:8], 16)
target = significand * (base ** exponent)

def spin(num_particles):
    maximumiter = math.ceil(30000000/num_particles)
    i = 0
    nonce = -1
    while i < 150000:
        for j in range(0, num_particles):
            nonce += 1
            header_hash = sha256(sha256(version + previous_header_hash + merkle_root + date + nbits + pack("<L", nonce)).digest()).digest()
            if int(hexlify(header_hash[::-1]), 16) < target:
                print("Nonce and Header Hash", nonce, hexlify(header_hash[::-1]))
                return
        i+=1
    return

if __name__ == "__main__":
    starttime = time.time()
    spin(num_particles=200)
    print("Time ", time.time() - starttime)

\end{lstlisting}

\end{document}